\documentclass[aip,reprint]{revtex4-1}

\usepackage[version=4]{mhchem}

\draft % marks overfull lines with a black rule on the right
\usepackage{graphicx}
\usepackage{subcaption}
\usepackage{xcolor}
\usepackage{ulem}
\usepackage{nicefrac}
\usepackage{comment}

\usepackage{hyperref}
\hypersetup{
    colorlinks=true,
    linkcolor=blue,
    filecolor=magenta,      
    urlcolor=cyan,
    citecolor=magenta
    }
    
\begin{document}

\title{High and magnetic-field-dependent surface carriers mobility in 3D topological insulators without bulk states} 

\author{M.V. Pugachev}
\affiliation{ 
P. N. Lebedev Physical Institute, Russian Academy of Sciences, Moscow 119991, Russia 
}
\author{A.E. Borisov}
\affiliation{ 
P. N. Lebedev Physical Institute, Russian Academy of Sciences, Moscow 119991, Russia 
}
\affiliation{ 
HSE University, Moscow, 101000, Russia
}
\author{A.V. Shupletsov}
\affiliation{ 
P. N. Lebedev Physical Institute, Russian Academy of Sciences, Moscow 119991, Russia 
}
\author{V.O. Sakhin}
\author{E.F. Kukovitsky}
\affiliation{ 
Zavoisky Physical-Technical Institute, FRC Kazan Scientific Center, Russian Academy of Sciences, 420029 Kazan, Russia
}
\author{A.Yu. Kuntsevich}
\email{alexkun@lebedev.ru}
\affiliation{ 
P. N. Lebedev Physical Institute, Russian Academy of Sciences, Moscow 119991, Russia 
}

\date{\today}

\begin{abstract}
By applying the conventional two-liquid model to the magnetoresistivity tensor, we reveal a record-high carrier mobility for surface states in tetradymite topological insulators ($\sim$ 20000 cm$^2$/Vs) in both bulk crystals and thin flakes of Sn-Bi${1.1}$Sb${0.9}$Te$_2$S. Bulk crystals of this 3D topological insulator exhibit a transition from bulk to surface-dominated conductivity below 100~K, whereas in thin flakes, bulk conductivity is suppressed at even higher temperatures. Our data therefore suggest that a key ingredient for elevated mobility is the absence of bulk carriers at the Fermi level. A fingerprint of the high-mobility carriers, i.e a steep low-field magnetoresistance along with a strong Hall effect nonlinearity below 1~T, signifies the presence of at least two surface-related carrier species, even when bulk states are frozen out. To explain the magnetoresistance and the Hall effect in a wider range of magnetic fields ($>1$~T), one must assume that the carrier mobility drops with the field. The influence of Zeeman splitting on mobility and the contribution of anomalous Hall conductivity provide a much better description of the magnetoresistance and the nonlinearity of the Hall coefficient. Our data call for a revision of the surface state mobility in 3D topological insulators.
\end{abstract}

\maketitle 

The absence of backscattering is a crucial characteristic of the topological surface states in 3D topological insulators (TIs), which implies high carrier mobility and a variety of potential applications\cite{Hasan2010, Tian2017}. In 3D TIs without magnetic impurities, backscattering is prevented by spin-momentum locking.
However, in practice, the mobility of the surface states is finite due to small-angle phonon or static interface disorder scattering. Moreover, due to surface depletion or accumulation, quantum well (or Rashba) states may form at the surface from the bulk bands. These states coexist with the topological ones\cite{zhu2011rashba,chen2012robustness, banerjee2016accessing, frantzeskakis2017trigger} and sometimes cannot be disentangled easily from them. In the most extensively studied family of 3D topological insulators, the (Bi,Sb)-chalcogenides with tetradymite structure, the highest reported surface states mobility in thin films is slightly over 10,000 cm$^2$/Vs \cite{Koirala2015, taylor2024magnetotransport}. Typically, mobilities around 5,000 cm$^2$/Vs are observed in samples where surface states coexist with bulk states at the Fermi level \cite{Moon2018, Park2016, Kuntsevich2016, Kim2014, Hoefer2014, Han2018}. The bulk carrier mobilities in the cutting-edge tetradymite TI single crystals were even higher --- up to several tens of thousands cm$^2$/Vs \cite{Bathon2016, Kokh2014, Paglione2010}. A fundamental and practical question therefore arises: why is the surface states mobility so low and how can it be increased?

In terms of surface states' carrier mobility, bulk states play a dual role: (i) bulk carriers screen the static disorder potential due to the elevated density of states at the Fermi level, which {\bf increases} the mobility of the surface states\cite{skinner}; (ii) scattering between the bulk and the surface states {\bf decreases} the mobility of the latter.

In most tetradymite 3D TIs, i.e., materials of the Bi$_2$Se$_3$ family, bulk states at the Fermi level arise due to unintentional doping through point defects \cite{Scanlon}. Considerable efforts have been made to shift the Fermi level into the bulk band gap, including methods such as doping \cite{Hong2012, Kushwaha2014}, partial substitution of the metal (Bi-Sb)\cite{Bai2020}, or chalcogen \cite{Ren2010}, and electrostatic gating \cite{Taskin2017}. In 2016, Kushwaha et al. suggested an optimal composition --- Sn-doped Bi${1.1}$Sb${0.9}$Te$_2$S (Sn-BSTS) \cite{Kushwaha2016}, which was subsequently reproduced by numerous research groups\cite{Cheng2016, Misawa2017, Wu2018, Cai2018, Liu2019, Sakhin2022, hattori2024validity, gudac2025unconventional}. This material demonstrates a Dirac-cone-like dispersion of surface states in the center of a 320~meV bulk band gap, with the Fermi level positioned within the band gap at temperatures $<100$ K. Even in bulk crystals of this material, 2D phenomena such as the quantum Hall effect \cite{Ichimura2019}, the planar Hall effect \cite{Wu2018} and 2D weak localization\cite{Mal2021} are observed. In the present paper, we focus on examining the carrier mobility in 3D TIs that have no bulk states at the Fermi level.

Carrier mobilities are typically determined using a two- or three-liquid model fit of magnetoresist{ivity} and Hall effect data. This model assumes the presence of parallel conductive channels, with their Drude conductivity tensors summed up\cite{Qu,Ren,Taskin}:
\begin{equation}
    \hat{\rho}^{-1}=\hat{\sigma}=\sum_i \frac{n_ie\mu_i}{1+(\mu_i B)^2}
    \begin{pmatrix}
        1 & -\mu_i B\\
        \mu_i B & 1
        \end{pmatrix},
    \label{multiliq}
\end{equation}
where $\hat{\rho}$ is the resistivity tensor, $\hat{\sigma}$ is the conductivity tensor, $\mu_i$ is the mobility of the $i$-th component, $n_i$ is the 2D density of the $i$-th component and $B$ is the magnetic field. The fitting parameters $\mu_i$ and $n_i$ are usually considered to be constants.

Numerous signatures indicate additional magnetoresistance mechanisms that superimpose on the few-liquid magnetoresistance in 3D TIs: (i) linear magnetoresistance, observed in Bi$_2$Te$_2$Se\cite{Assaf2013}, Bi$_2$Te$_3$\cite{Wang2015, Shrestha2017}, BiSbTeSe$_2$\cite{taylor2024magnetotransport,zhu2025improvement}, suggests a mobility drop with magnetic field; (ii) an anomalously large prefactor in weak antilocalization, observed in 3D TI thin films \cite{Chen2011, Assaf2013, Stepina2023}, cannot be explained by quantum interference alone; and (iii) low-field ($< 1$~T) Hall effect nonlinearity, observed in Refs.\cite{Stepina2023} and \cite{Shrestha2017} at magnetic fields much smaller than the inverse mobility, might be an indicator of either high-$\mu$ carriers or an additional Hall effect contribution mechanism.

In the present study, we carefully measure the resistivity and Hall resistivity of the model systems, i.e., Sn-BSTS crystals and thin flakes in a wide range of temperatures. The similarity of the results obtained on crystals and thin flakes at low temperatures clearly identifies the surface nature of the observed features. We find that constant values of $\mu_i$ fail to describe the data. Low-field Hall effect non-linearity and steep magnetoresistance are indicative of a high-mobility group of carriers. The mobility tends to increase as the temperature decreases, exceeding 20,000 cm$^2$/Vs below 50 K.

However, data at higher magnetic fields cannot be described by high-mobility carriers, indicating a decrease in mobility with increasing field. We discuss why the elevated mobility and its field dependence were overlooked in previous studies and consider possible mechanisms for the field-dependent mobility.

The crystals with nominal composition Sn$_{0.02}$Bi$_{1.08}$Sb$_{0.9}$Te$_2$S were grown using the Bridgman method from the melt and characterized as described in Ref.\cite{Sakhin2022}. These crystals had a mirror-like cleavage plane and exfoliated easily. Besides bulk crystals with a thickness of about 70 $\mu$m, shaped as a rectangular bar (an optical micrograph is shown in Fig.~\ref{fig:R_from_T}\textbf{a}), we also mechanically exfoliated 3.2~$\mu$m and 200~nm thick flakes (a scanning electron microscope image is shown in Fig.~\ref{fig:R_from_T}\textbf{b}) onto an SiO$_2$/Si substrate. E-beam evaporated Ti/Al contacts were fabricated using shadow mask lithography\cite{Micromask} and lift-off techniques. The Hall-bar mesa was defined using gallium ion beam etching. Hall bar geometry is necessary for quantitative determination of the conductivity tensor components.

The temperature dependencies of the resistivity per square for all three samples are shown in Fig.~\ref{fig:R_from_T}\textbf{c}, and they agree with those reported by other groups for bulk Sn-BSTS crystals\cite{Misawa2017, Wu2018, Cai2018, Liu2019, Sakhin2022, Zhao2019} and their thin flakes\cite{Ichimura2019, Misawa2020, Matsushita2021}. As the temperature decreases from 300~K, the bulk conductivity is frozen out, leading to a negative $d\rho_{xx}/dT$. At low temperatures (approximately 100 K and below), only surface conductivity remains, with $\rho_{xx}\sim 1$~k$\Omega$ and a positive $d\rho_{xx}/dT$ due to electron-phonon scattering of metallic-like surface carriers for all sample thicknesses. The low-temperature resistivity values and temperature behavior differ between samples because they were exfoliated from different parts of the bulk crystal\cite{Kushwaha2016}. The 200~nm flake experienced additional photoresist and Ga ion beam exposure, leading to doping and changes in static disorder; therefore, its low-temperature behavior deviates more strongly. The magnetic field was swept from positive to negative values at a rate of 0.1~T/min. Magnetoresistance (Hall resistance) dependencies were measured and symmetrized (antisymmetrized) to compensate for the inevitable contact misalignment. Hereafter, we consider the system as two-dimensional and calculate its resistivity using the sample width and the distance between voltage contacts. The Hall resistivity is equivalent to the Hall resistance. Both the resistivity and Hall resistivity are measured in Ohms. The current density did not exceed 10~$\mu$A/mm to avoid overheating.

\begin{figure}
    \centering
    %\resizebox{0.5\textwidth}{!}{\includegraphics{1.R(T).png}}
    \includegraphics[width=0.48\textwidth]{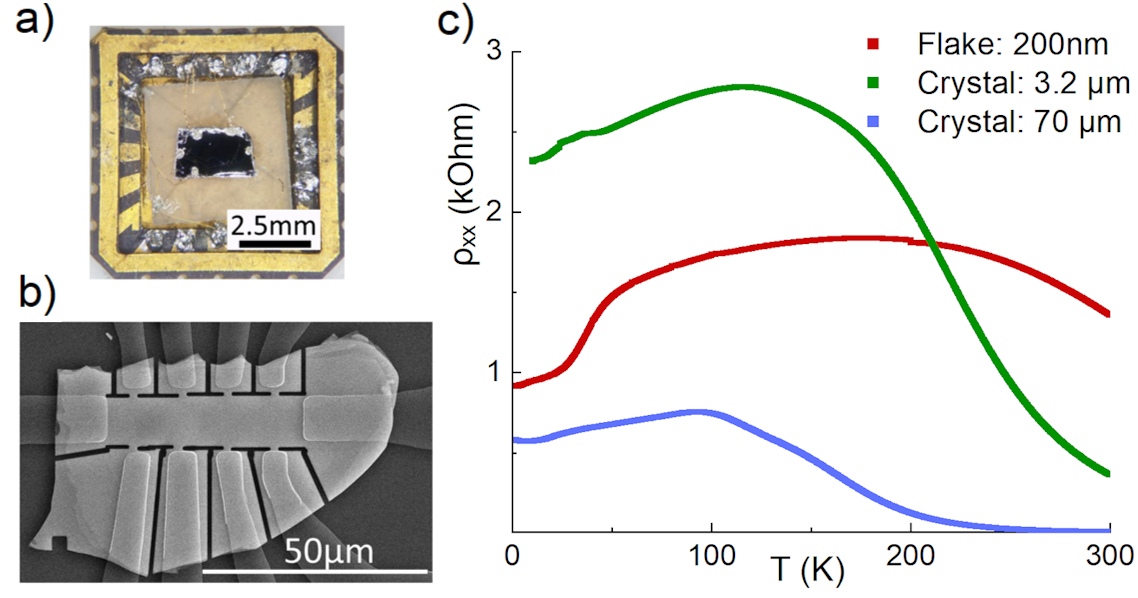}
    \caption{\textbf{(a)} Optical micrograph of the bulk sample mounted for measurements with silver-paint-glued contacts. \textbf{(b)} Scanning electron microscope image of the flake with contacts. \textbf{(c)} Temperature dependencies of the resistivity per square for the flake (red) and bulk crystal samples (blue and green).}
    \label{fig:R_from_T}
\end{figure}

Fig.~\ref{fig:R_from_B}\textbf{a}--\textbf{c} shows the longitudinal resistivity $\rho_{xx}$, transverse resistivity $\rho_{xy}$, and Hall coefficient $\rho_{xy}/B$ dependencies on magnetic field $B$ at different temperatures for the 200~nm flake. Similar results for the bulk and intermediate-thickness samples are shown in Fig.~S1 in the Supplementary Information. At high temperatures, the sign of the Hall coefficient corresponds to p-type in the bulk crystal (see Fig.~S1\textbf{b,c} in the Supplementary Information) and n-type in the flake. At low temperatures, the Hall effect in both the flake and bulk samples exhibits an n-type sign. These observations are typical for this material\cite{Ichimura2019} and indicate downward band bending near the surface. Positive magnetoresistance and the field-dependent Hall coefficient suggest the presence of at least two types of carriers in Sn-BSTS. We rule out a difference between the top and bottom surfaces because the effects are observed in bulk crystals, where these surfaces are certainly equivalent. We speculate that, in addition to topological surface states, the other group of carriers originates from near-surface quantum well states.

\begin{figure*}
    \centering
    %\resizebox{0.9\textwidth}{!}{\includegraphics{2.R(B).png}}
    \includegraphics[width=0.92\textwidth]{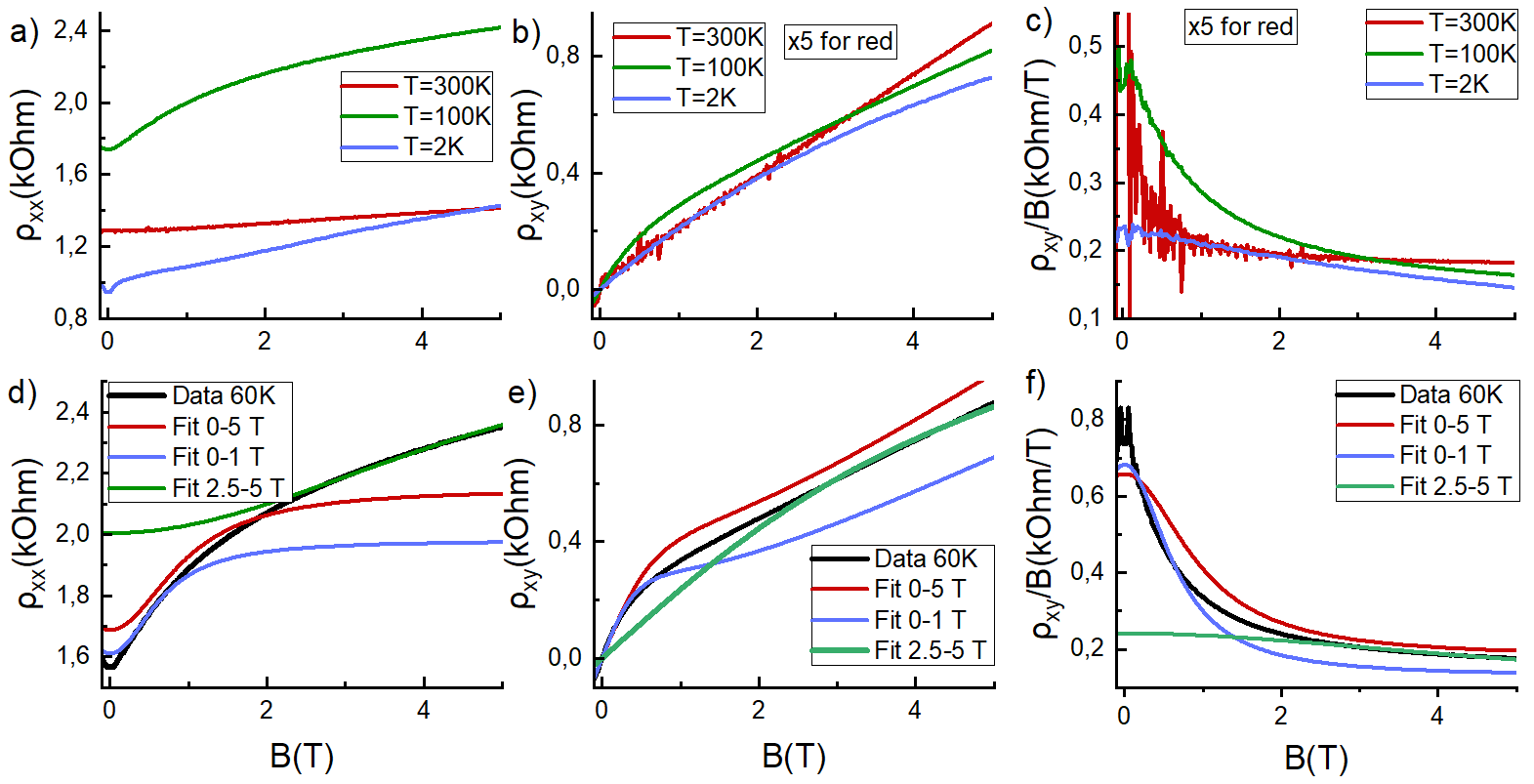}
    \caption{Representative magnetoresistivity $\rho_{xx}(B)$ (\textbf{a}), Hall resistivity $\rho_{xy}(B)$ (\textbf{b}) and Hall coefficient $\rho_{xy}/B(B)$ (\textbf{c}) experimental data for the flake. Panels (\textbf{d}--\textbf{f}) show a comparison of the $\rho_{xx}(B)$, $\rho_{xy}(B)$, and $\rho_{xy}/B(B)$ data at 60~K (black curves), respectively, with two-liquid fits performed in various magnetic field ranges (color curves).}
    \label{fig:R_from_B}
\end{figure*}

The most pronounced $\rho_{xx}(B)$ and $\rho_{xy}(B)$ nonlinearities are observed at $T\approx60$~K, as shown in Fig.~\ref{fig:R_from_B}\textbf{d}--\textbf{f}. The effect on the Hall coefficient may be especially dramatic. In Fig.~\ref{fig:R_from_B}\textbf{f}, $\rho_{xy}/B$ drops from 800~$\Omega$/T to 400~$\Omega$/T as the magnetic field increases from 0 to 1~T. 

Fitting the data with a two-carrier model (Eq.~\ref{multiliq}) requires the use of the least squares method with four adjustable parameters ($\mu_1,n_1, \mu_2, n_2$). Apparently, the simultaneous fit of both $\rho_{xx}(B)$ and $\rho_{xy}(B)$ with the two-liquid model is misleading, because at low magnetic fields $\rho_{xy}$ tends to zero, and the Hall coefficient feature does not contribute significantly to the mean square deviation from Eq.~\ref{multiliq}. To address this issue, we minimize the functional:
\begin{equation}
    \Delta f = \sum_i [(\rho_{xx}^{ex}(B_i)-\rho_{xx}^{th}(B_i)^2 (\frac{\rho_{xy}^{ex}(B_i) - \rho_{xy}^{th}(B_i)}{B_i})^2].
\end{equation}

Here $i$ is the index of data points (corresponding to different magnetic field values), superscripts ``$ex$'' and ``$th$'' denote the experimental data and the result of calculation using Eq.~\ref{multiliq}, respectively, and $B_{r}$ is an effective magnetic field that defines the weight of the Hall coefficient data in the fit. The value of $B_{r}$ was adjusted to ensure that the contributions of $\rho_{xx}$ and $\rho_{xy}/B$ to $\Delta f$ are approximately equal.

It turns out that the data can only be well-fitted within limited ranges of magnetic fields. For example, if only low magnetic field data points below 1~T are used for fitting (the green line in Fig.~\ref{fig:R_from_B}\textbf{d}--\textbf{f}), significant deviations from the fit occur at higher fields. The red line corresponds to the least-square criterion for the whole range of fields (up to 5~T), but it deviates at both high and very low fields.

The blue curve fits the region above 2.5~T well but deviates significantly at low fields. The fitting parameters (densities and mobilities) for the low-field and high-field fits vary significantly. The low-field mobility for one group of carriers is approximately 20000~cm$^2$/Vs, whereas at high magnetic fields, the mobility is 5-8 times lower. The extraordinarily high mobility in a 3D TI without bulk conductivity, extracted using a conventional two-liquid magnetoresistance fit in the low-field regime, and the mobility drop at higher fields are the major experimental observations of our paper.

Using a three-liquid fit does not allow us to describe the magnetoresistivity tensor over the whole range of fields. Any Drude tensor model produces the nonlinearity onset in both magnetoresistivity and the Hall coefficient for $\mu B\sim 1$. In our experiments, this field scale is below $0.5$~T (see Fig.~\ref{fig:R_from_B}\textbf{d}--\textbf{f}). At the same time, the high temperature rules out a weak antilocalization (WAL) contribution: a small WAL correction with an amplitude on the order of $e^2/h$ is indeed observed at low temperatures.

Fig.~\ref{fig:n_mu_flake} shows the temperature dependencies of the densities and mobilities derived from the simultaneous magnetoresistivity and Hall coefficient two-carrier model fit. The figure compares three ranges of magnetic fields used for fitting. For temperatures below 35~K, data for magnetic fields $<$~0.4~T were disregarded to eliminate the WAL contribution. We emphasize that the field dependence of the fitted parameters means that the two-liquid model provides only rough estimates and must be revised.

\begin{figure}[!ht]
    \centering
    %\resizebox{0.45\textwidth}{!}{\includegraphics{3.4-pars_fit.png}}
    \includegraphics[width=0.48\textwidth]{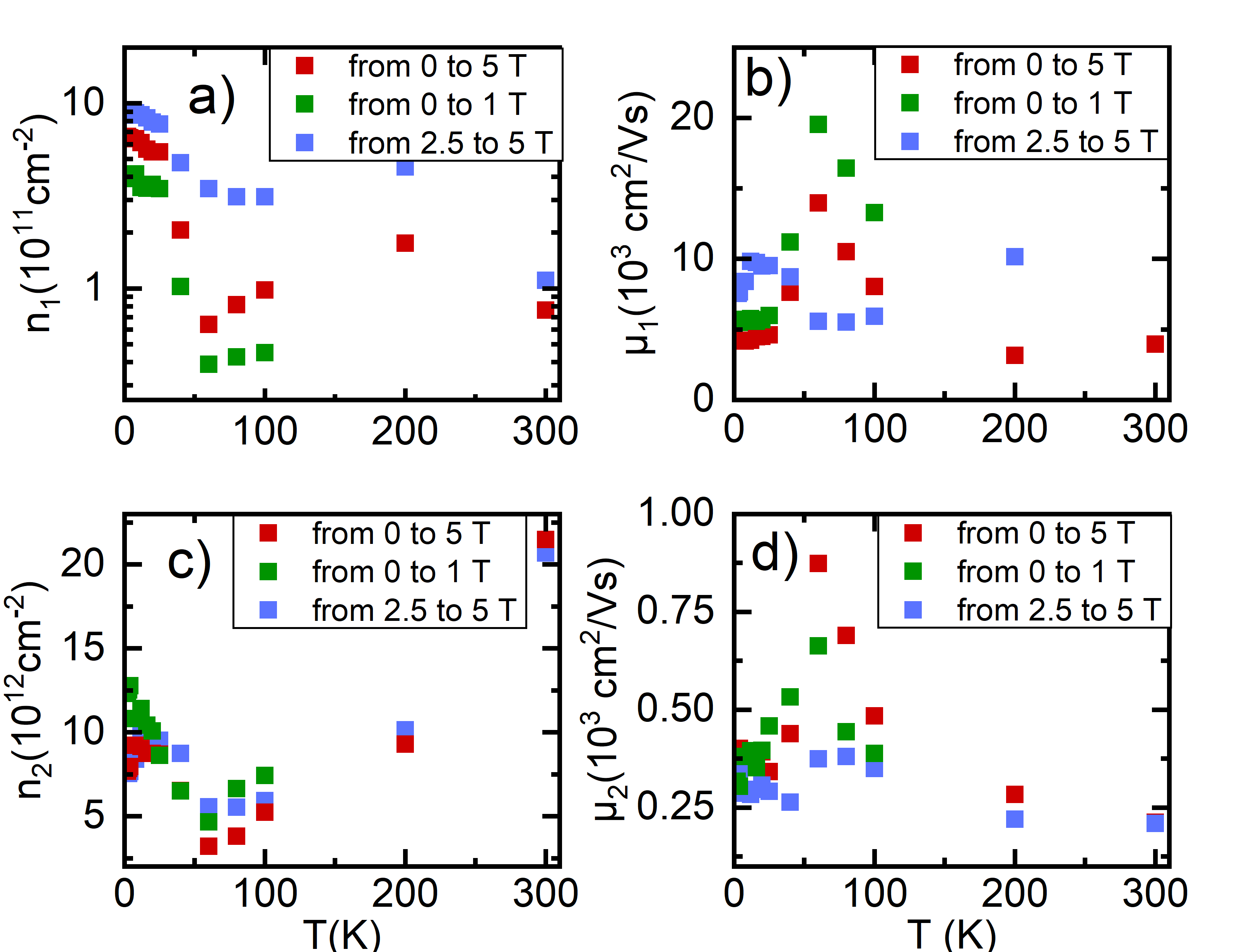}
    \caption{Temperature dependence of the carrier densities per sheet ($n$; panels \textbf{a} and \textbf{c}) and mobilities ($\mu$; panels \textbf{b} and \textbf{d}) for the Sn-BSTS flake, obtained from the simultaneous fit of $\rho_{xx}(B)$ and $\rho_{xy}(B)$ dependencies with the two-liquid model, as explained in the text. The top panels (\textbf{a} and \textbf{b}) and bottom panels (\textbf{c} and \textbf{d}) show the values for the high- and low-mobility components, respectively.}
    \label{fig:n_mu_flake}
\end{figure}

The two-liquid model does not specify the nature of the fit parameters (densities $n$ and mobilities $\mu$). The carriers of the first type have the highest mobility. Their $T$-dependencies of $n_1$ and $\mu_1$ are shown in Figs.~\ref{fig:n_mu_flake}\textbf{a} and \textbf{b}, respectively. We believe they are topological surface states. Their mobility depends strongly on the magnetic field range used for the fit. Regardless of the fitting range, the $n_2$ values (see Fig.~\ref{fig:R_from_T}\textbf{c}) clearly demonstrate growth with temperature above 100 K, signifying their bulk nature. Well below the temperature of the $\rho_{xx}(T)$ maximum, the bulk carriers are frozen out, and $n_2$ must correspond to the density of some other surface-related carriers. The mobilities of both groups of carriers tend to decrease at elevated temperatures due to phonon and activated carrier scattering.

The low-temperature ($<100$K) behavior could be interpreted as arising from two types of topological surface carriers from the bottom and top surfaces. However, quantitatively similar results for the bulk crystal, where the top and bottom surfaces are equivalent (see Fig.~S2 in the Supplementary Information), suggest a different origin for the second group of carriers. We attribute them to states of the bulk conduction band localized in the self-consistent potential due to near-surface band bending. These states are well-known in 3D TIs and are called quantum well or Rashba states\cite{chen2012robustness, zhu2011rashba, banerjee2016accessing, frantzeskakis2017trigger}. However, it is still difficult to explain the unphysical growth of $n_1$ by more than an order of magnitude at low temperatures. Therefore, some other magnetotransport mechanism must exist in addition to the Drude multi-component conductivity.

Previous experiments on Sn-BSTS show traces of the same phenomena, i.e. steep magnetoresistance and a nonlinear Hall effect beyond the multi-liquid Drude model. For example, steep magnetoresistance is seen in Ref.\cite{Kushwaha2016} (Fig.~5 therein). However, that study was focused on a too wide range of magnetic fields (up to 18~T). In Ref.~\cite{Wu2018} (Fig.~2\textbf{a} therein), the magnetoresistance curves are very similar to those shown above in Fig.~\ref{fig:R_from_B}\textbf{a}, although the Hall effect is not analyzed. Data with Hall nonlinearity are presented in Ref.\cite{Misawa2020} without multi-liquid analysis. Our data, together with signatures of an unconventional magnetoresistivity tensor in other 3D TIs\cite{Chen2011, Assaf2013,Wang2015, zhu2025improvement, Shrestha2017, Stepina2023}, suggest a need to revise the magnetotransport in 3D TIs and to consider the magnetic field dependence of mobility.

The observation of the record mobility in BSTS, where itinerant bulk carriers are almost absent, means that scattering between surface and bulk states is probably an important momentum relaxation mechanism in 3D TIs. A field-dependent mobility model must exploit the time-reversal symmetry breaking by the magnetic field. For example, the magnetic field could create a nonzero matrix element for backscattering, or the Zeeman component could open a gap in the surface states' spectrum, thus reducing the group velocity and, hence, the mobility. One should also possibly take into account an anomalous Hall effect due to spin-orbit interaction, which adds to the conventional one.

We suggest a minimal model where, besides the Lorentz force (which curves electron trajectories and leads to Eq.~\ref{multiliq}), the mobility of one group of carriers becomes field{-}dependent. The physics is related to a Zeeman term $g\mu_B\sigma_zB_z$ in the Hamiltonian: $\hat{H}=s (p_x \sigma_y-p_y\sigma_z) +g\mu_B\sigma_zB_z$. Here the first two terms are the standard TI surface states' linear-in-momentum spectrum and $\sigma_i$ are the Pauli matrices. The Zeeman term leads to gap opening and the appearance of otherwise prohibited backscattering. This spectrum is similar to that of a magnetic or a magnetically proximized topological insulator, as considered theoretically in Refs.\cite{chiba, akzyanov2018}. For a single surface, it leads to the following variation in the conductivity tensor\cite{chiba}:

\begin{equation}
    \sigma_{xx}=\frac{\sigma_D}{1+4\alpha^2B^2}
    \label{sigmaxxmodel}
\end{equation}

\begin{equation}
  \sigma_{xy}=f_\alpha(B)=\frac{e^2}{h}\,\alpha B\,
  \sqrt{1+(\alpha B)^2}\,
  \frac{4\!\left[1+2(\alpha B)^2\right]}{\left[1+4(\alpha B)^2\right]^2}\,,
  \label{AHE}
\end{equation}
where $\alpha$ it the Zeeman effect parameter $g\mu_B/E_F$. The absence of an orbital contribution suggests to use Eq.~\ref{sigmaxxmodel} for the surface states mobility: $\mu_1(B)=\mu_1^0/(1+4\alpha^2B^2)$, where $\mu_1^0$ is the Drude value of the mobility at $B=0$. 

We fit Hall coefficient and resistivity with a two-liquid model with $B$-dependent $\mu_1$ value, i.e. $$\sigma_{xx}= \frac{n_1e\mu_1(B)}{1+{\mu_1(B)}^2B^2}+\frac{n_2e\mu_2}{1+{\mu_2}^2B^2},$$  The Hall conductivity tensor component contains an admixture of the anomalous component (Eq.~\ref{AHE}):
\[
\sigma_{xy}=\frac{n_1 e \mu_1(B)^2 B}{1+\mu_1(B)^2 B^2}
+\frac{n_2 e\,\mu_2^2 B}{1+\mu_2^2 B^2}
+ f_\alpha(B),
\]

Conductivity tensor inversion allows us to fit the experimental data with five adjustable parameters: $n_1$, $n_2$, $\mu_1^0$, $\mu_2$, $\alpha$. A detailed theoretical description is presented in Supplementary Information Section S3.

An example of the 5-parameter fit of magnetoresistivity and the Hall coefficient simultaneously across the entire magnetic field range is shown in Fig.~\ref{fig:R_from_B1}\textbf{a} and \textbf{b}. It is evident that this fit describes the data much better than the standard 4-parameter fit.

\begin{figure}[t]
    \centering
    %\resizebox{0.45\textwidth}{!}{\includegraphics{4.5-pars_fit.png}}
    \includegraphics[width=0.48\textwidth]{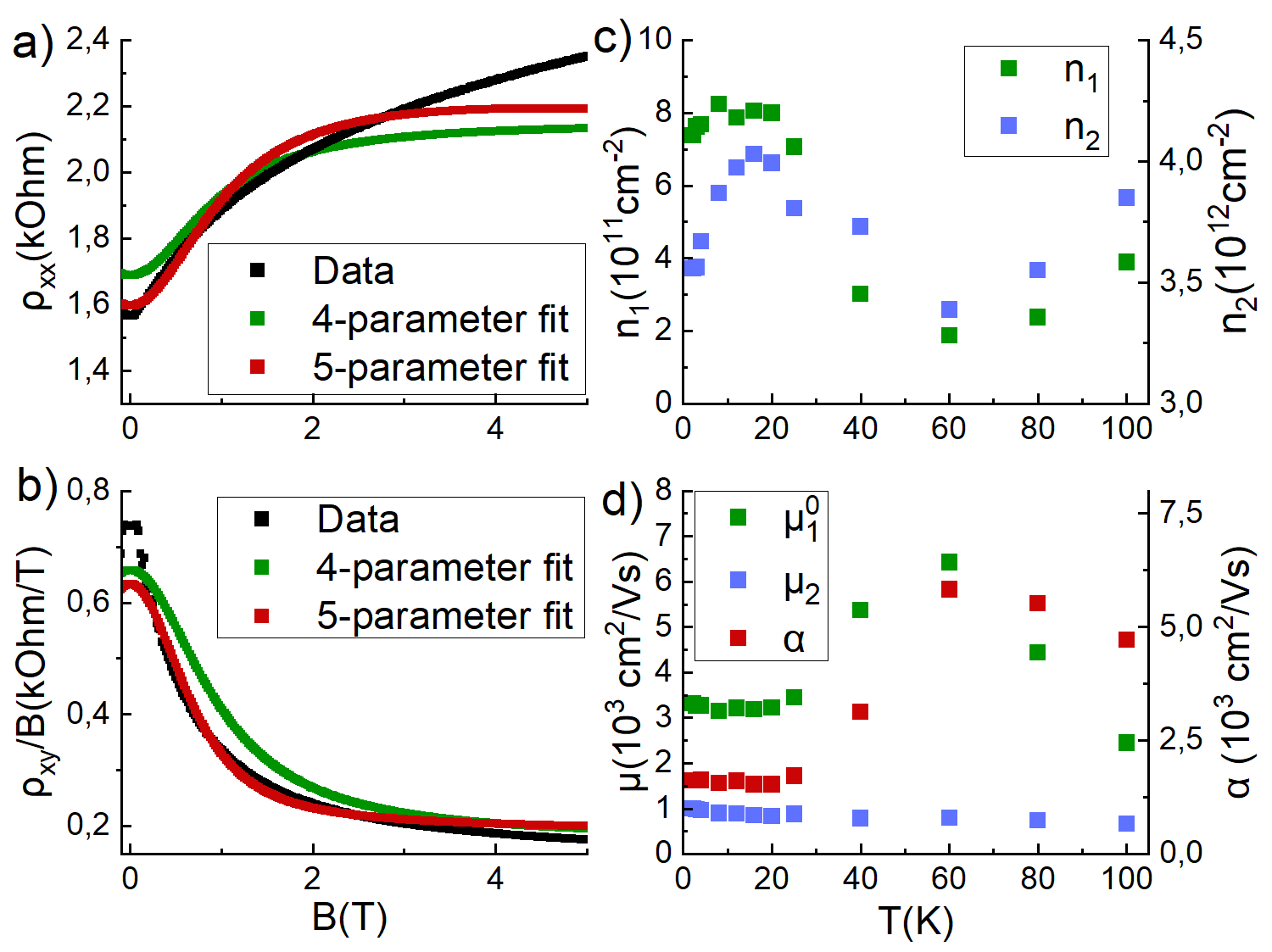}
    \caption{Magnetoresistivity (\textbf{a}) and Hall coefficient (\textbf{b}) for the flake at 60~K. Black curves are experimental data, red and green curves are 4- and 5-parameter fits, respectively. Panels (\textbf{c}) and (\textbf{d}) show temperature dependencies of the carrier density and mobility within the 5-parameter fit, respectively.}
    \label{fig:R_from_B1}
\end{figure}

The fit parameters (shown in Figs.~\ref{fig:R_from_B1}\textbf{c} and \textbf{d}) become more physical compared to the two-liquid four-parameter fit. The temperature dependencies of both the density $n_2$ and mobility $\mu_2$ are significantly suppressed. This observation signifies that there is a large reservoir ($\sim 3 \times 10^{12}$ cm$^{-2}$) of low-mobility carriers responsible for the high-field conductivity tensor. The peak mobility of the ``high-$\mu$'' group becomes smaller, about 6500 cm$^2$/Vs. This is understandable because a certain part of the magnetoresistance, compared to the 4-parameter model, now comes from the anomalous Hall effect contribution and the $\mu(B)$ dependence, i.e., from the $\alpha B$ factor. These additional mechanisms enhance the magnetoresistance and Hall effect nonlinearity, so high mobility values are no longer required. The unphysical $n_1(T)$ dependence becomes weaker (about a factor of 4 variation) compared to a factor of 10 in Fig.~\ref{fig:n_mu_flake}. This temperature dependence is the most puzzling observation in our paper.

The values of $\alpha$ are in the range from 2000 to 5000~cm$^2$/Vs, which corresponds to the onset of out-of-plane spin polarization (where the Fermi energy is comparable to the Zeeman energy) in magnetic fields of a few T. This is indeed possible because the $g$-factors are very high in 3D TI materials. Physically, this would mean the observation of the quantized anomalous Hall effect (QAHE) in a field of a few T. The transport observation of this phenomenon is hindered by the second group of carriers with high density and low mobility. Still, it is possible that the QAHE physics is responsible for the puzzling temperature dependence of $n_1$.

We believe that by proper adjustment of the field-dependent mobility model, one can achieve better agreement. Even within the model of Ref.\cite{chiba}, the results for fixed 2D density and fixed chemical potential are different. It is not known for sure which is the case in 3D TIs\cite{kuntsevich2018}. The mobility of the Rashba states could also be field-dependendent\cite{wang2022large}. Possible non-uniformity of the system, leading to positive magnetoresistance\cite{Parish,Kuntsevich2020}, is neglected here. All these factors are hard to take into account. Two facts remain firmly established by this research: high mobility and its drop with field.

An indirect conclusion from our data is that at sufficiently high magnetic fields, the mobility is rather low. This inference is in agreement with previous observations of the Shubnikov-de Haas oscillations onset for $B \sim 10$~T\cite{Kushwaha2016, gudac2025unconventional}, which corresponds to a mobility of about $1/B \sim 1000$~cm$^2$/Vs. Indeed, at low magnetic fields, the mobility is high, leading to a nonlinear Hall effect and magnetoresistance. However, at higher fields, the mobility drops, and the system becomes effectively more disordered, which, in turn, suppresses quantum oscillations.

In conclusion, we reveal that the magnetoresistivity and Hall coefficient behavior for Bi$_{1.1}$Sb$_{0.9}$Te$_2$S-Sn topological insulator single crystals and thin flakes is qualitatively similar to that of the two-liquid model, but with high and magnetic-field-dependent mobility. The highest effective mobility values, which are record-setting for tetradymite topological insulators, are observed below 1~T. Our data suggest that the absence of bulk states could promote the elevated mobility. Similar effects also emerge in the other 3D topological insulator materials. One could explain the observed effects by taking into account Zeeman splitting, which breaks time-reversal symmetry. Our data thus calls for (i) an experimental revision of the mobility in topological insulators (the mobilities must rise in the low-magnetic-field limit) and (ii) the development of a field-dependent mobility theory.

\textbf{Appendix.} Section S1 presents the magnetotransport data for bulk crystals and their fits; Fig.~S1\textbf{a}--\textbf{c} is the analog of Fig.~\ref{fig:R_from_B}\textbf{a}--\textbf{c} for bulk crystals; Fig.~S2 is an analog of Fig.~\ref{fig:n_mu_flake}. Supplementary Section S2 with Fig.~S3 compares the results of the conventional 4-parameter and Zeeman-included 5-parameter models. Section S3 contains the theoretical description of the magnetic field dependence of the mobility. Fig.~S4 shows the magnetotransport models discussed in the paper and their interrelations.

\textbf{Acknowledgements.} The authors are thankful to A.L.~Rakhmanov and R.S.~Akzyanov for stimulating discussions, and G.B.~Teitelbaum for simulating the crystal growth in Zavoisky Physical-Technical Institute. The work is supported by Russian Science Foundation Grant 25-42-01036. The work of E.F.K. concerning the crystal growth was funded by the Russian Science Foundation according to Research Project No. 21-72-20153-P. The work of V.O.S. concerning the initial crystal characterization was supported by the government assignment for FRC Kazan Scientific Center of Russian Academy of Science.The measurements and microfabrication were performed at the Shared Facility center at the P.N.~Lebedev Physical Institute.

\bibliography{biblio}
\newpage
\renewcommand{\thepage}{S\arabic{page}}
\renewcommand{\thesection}{S\arabic{section}}
\renewcommand{\thetable}{S\arabic{table}}
\renewcommand{\theequation}{S\arabic{equation}}
\renewcommand{\thefigure}{S\arabic{figure}}

\appendix

\maketitle
\end{comment}

\section{Data for bulk samples}

This section demonstrates that the bulk samples do not differ from the thin flake (200~nm) in terms of the physical effects discussed in the paper.

\begin{figure}[!htbp]
  \centering
  \includegraphics[width=\linewidth]{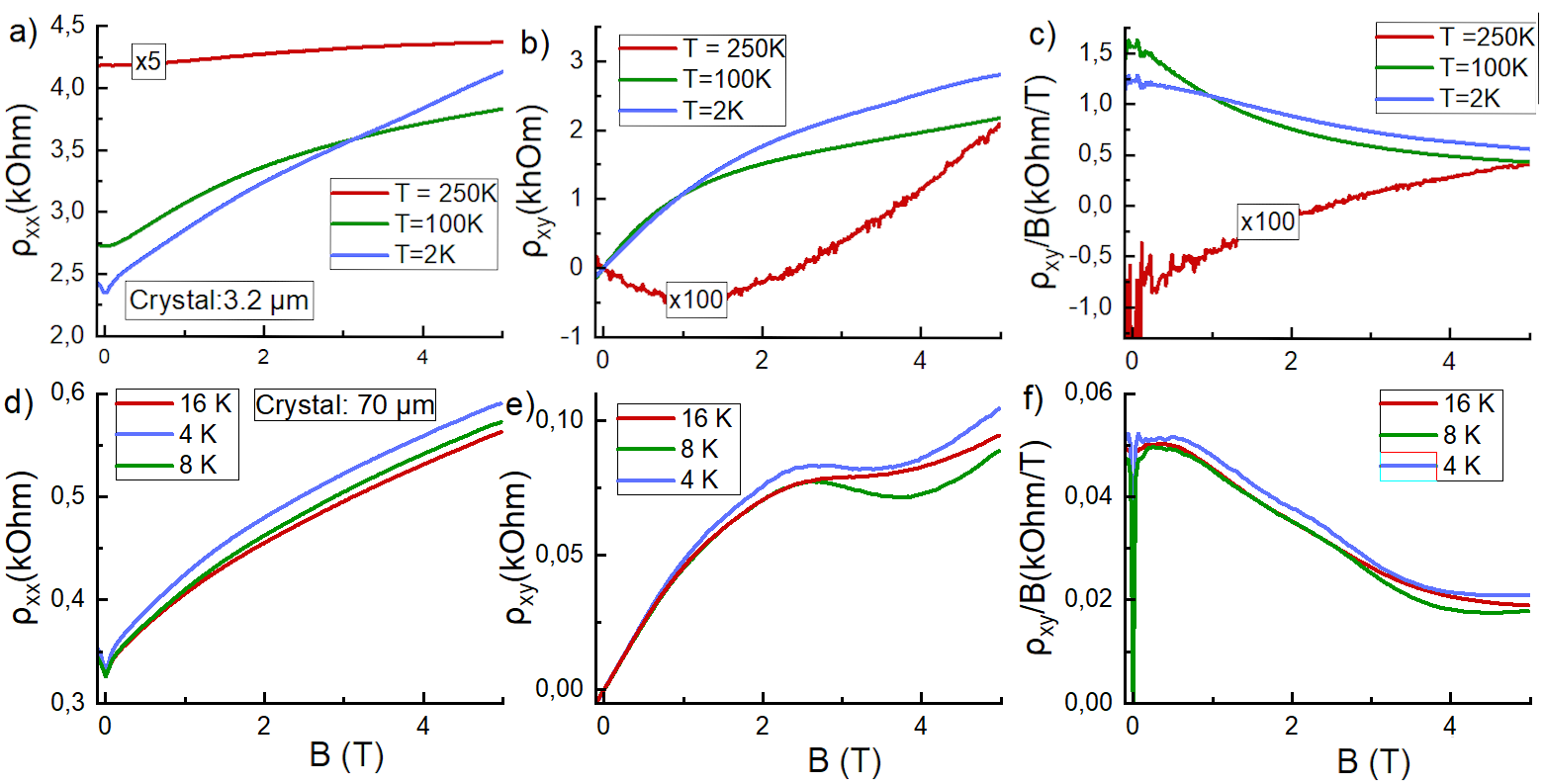}
  \caption{Representative magnetoresistivity $\rho_{xx}(B)$ (panels \textbf{a} and \textbf{c}), Hall resistivity $\rho_{xy}(B)$ (panels \textbf{b} and \textbf{e}) and Hall coefficient $\rho_{xy}(B)/B$ (panels \textbf{c} and \textbf{f}) experimental data for the crystals. Panels (\textbf{a}–\textbf{c}) correspond to the 3.2~$\mu$m-thick crystal; panels (\textbf{d}–\textbf{f}) correspond to the 70~$\mu$m-thick crystal.}
  \label{fig:sup:R_B_bulk}
\end{figure}

Fig.~\ref{fig:sup:R_B_bulk} shows the magnetic field dependencies of the resistivity $\rho_{xx}$, Hall resistivity $\rho_{xy}$, and Hall coefficient $\rho_{xy}/B$ for bulk samples with thicknesses of 3.2~$\mu$m and 70~$\mu$m. The magnetoresistivity is smooth and positive, similar to the curves in Fig.~2\textbf{a} of the main text (for the flake). At low temperatures, the Hall coefficient has a peak at zero field and a smooth behavior, also similar to the curves in Fig.~2\textbf{c} of the main text. At high temperature (250~K), a non-monotonic behavior with a sign change is observed for the Hall resistivity of the 3.2~$\mu$m-thick bulk sample. This is caused by the dominant contribution of bulk holes compared to that of surface electrons. In the flake, the role of bulk holes is suppressed; therefore, the sign of the Hall resistivity is preserved.

\begin{figure}[t]
  \centering
  \includegraphics[width=\linewidth]{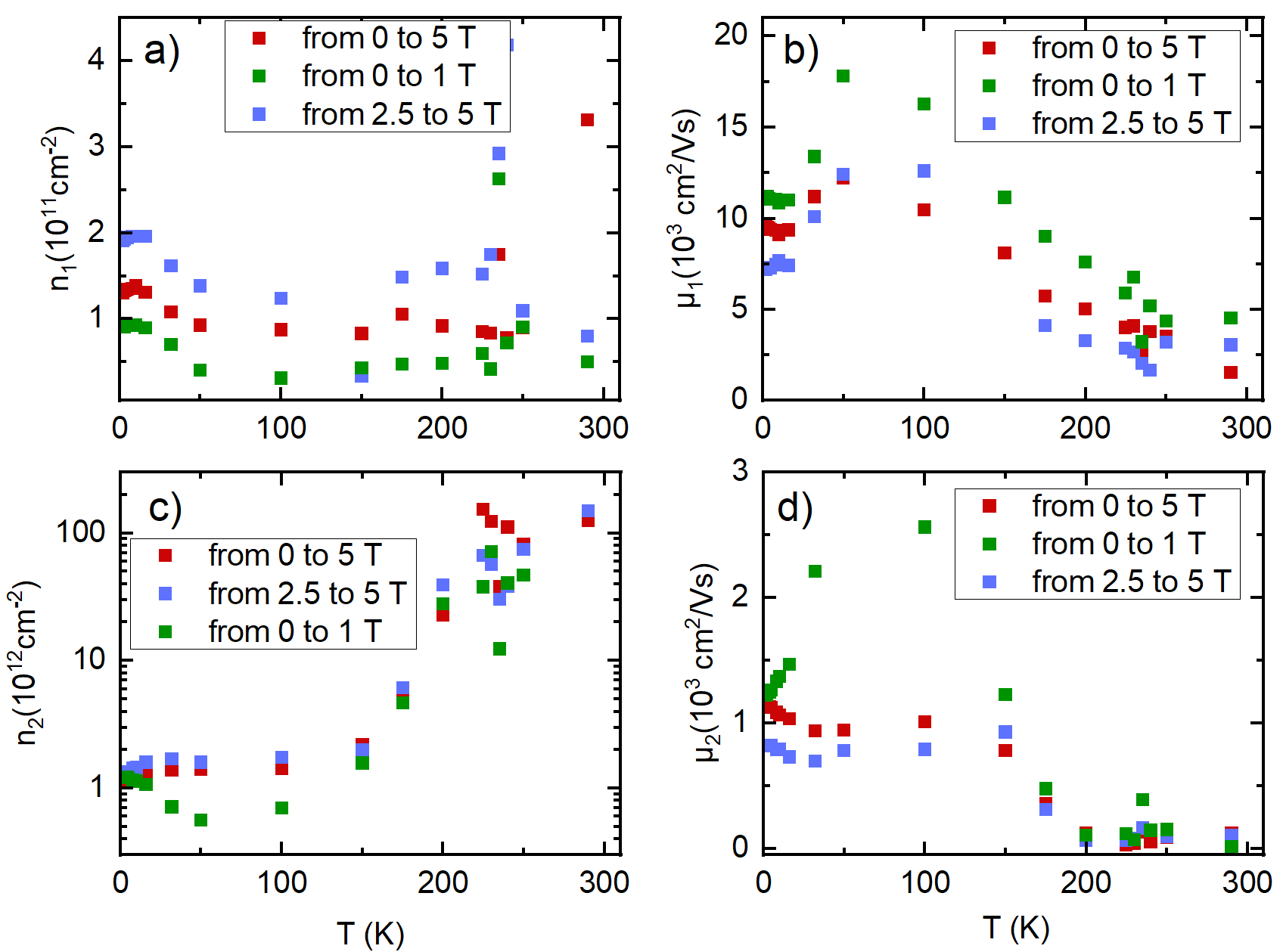}
  \caption{{Temperature dependence of the sheet carrier densities ($n$; panels \textbf{a} and \textbf{c}) and mobilities ($\mu$; panels \textbf{b} and \textbf{d}) for a 3.2\,$\mu$m-thick Sn--BSTS bulk crystal, obtained from simultaneous fits to $\rho_{xx}(B)$ and $\rho_{xy}(B)$ using the two-carrier model. The top panels (\textbf{a}, \textbf{b}) show the high-mobility component; the bottom panels (\textbf{c}, \textbf{d}) show the low-mobility component.}}
  \label{fig:sup:n_mu_bulk}
\end{figure}

Fig.~\ref{fig:sup:n_mu_bulk} shows the parameters obtained from the two-liquid model fit of the data for the 3.2~$\mu$m-thick bulk sample. The quantitative behavior of the parameters is similar to that in Fig.~3 of the main text. The high-mobility component ($\mu_1$ and $n_1$) exhibits a peak in mobility and a drop in density near 60~K. The values of the extracted mobility are of the same order and systematically higher for the low-field fit. The low-mobility component shows a peak in mobility and the onset of a plateau in density near $T \sim 100$~K. At high temperatures, the density increases significantly and the mobility decreases with increasing temperature. We wish to point out that the scale of $n_2$ at high temperatures is about an order of magnitude higher than that for the flake (see panel Fig.~3\textbf{c} in the main text).
%\clearpage

\section{Comparison of the models results}

\begin{figure}[!htbp]
  \centering
  \includegraphics[width=\linewidth]{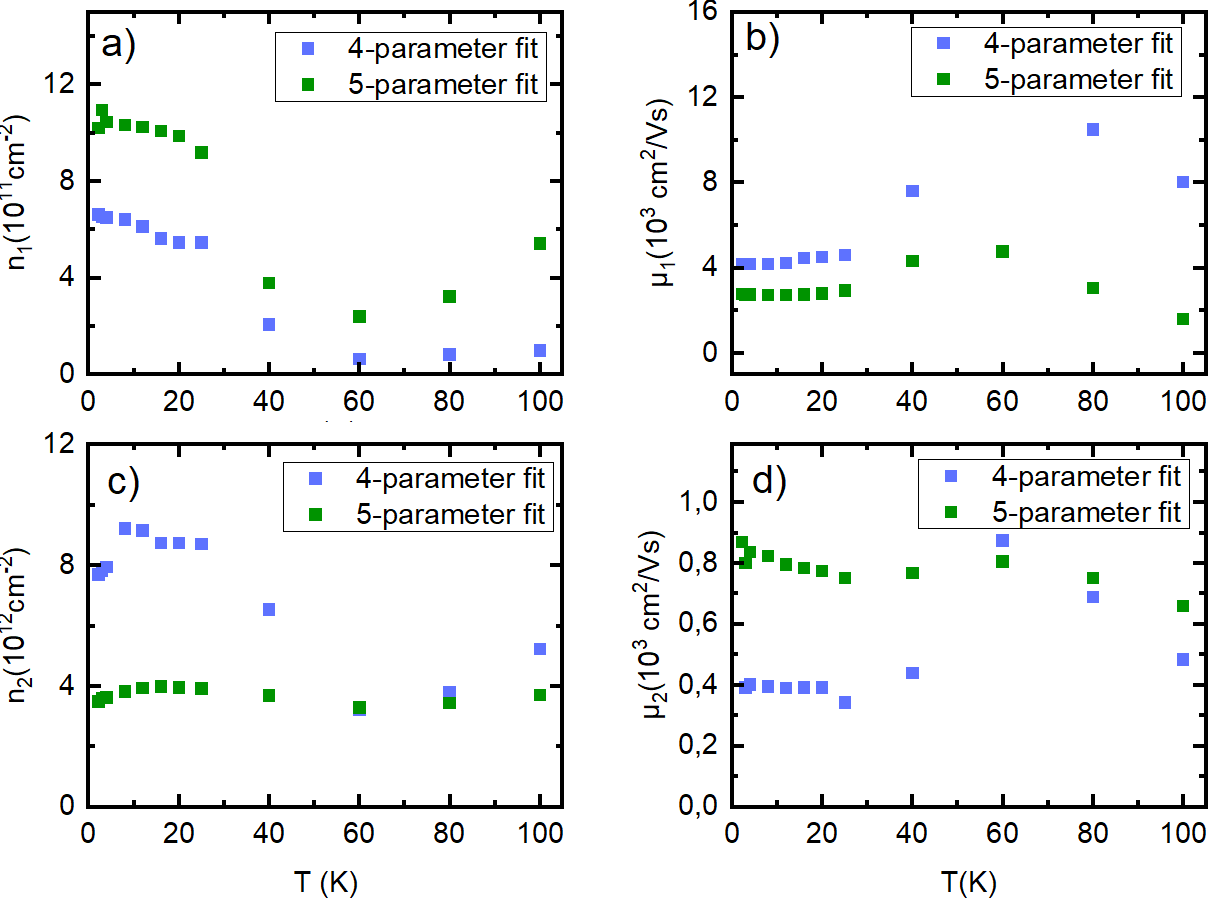}
  \caption{{Comparison of carrier densities ($n$, panels \textbf{a} and \textbf{c}) and mobilities ($\mu$, panels \textbf{b} and \textbf{d}) obtained from the 4-parameter and 5-parameter fits for the flake (200~nm thickness) at various temperatures.}}
  \label{fig:sup:mod_comp}
\end{figure}

Fig.~\ref{fig:sup:mod_comp} shows a comparison of the parameters $n_1$, $n_2$, $\mu_1$, and $\mu_2$ obtained from the 4-parameter (conventional) and 5-parameter (with $\mu(B)$ dependence included) models. For the 5-parameter model, panel \textbf{b} shows the values of $\mu_1^0$. The upgraded model considers the magnetic field dependence of only the high-mobility component ($\mu_1$). The 5-parameter model gives systematically lower values for $\mu_1$ and higher values (except at $T=60$~K) for $\mu_2$. The densities behave in the opposite way: the density of the low-mobility component decreases, while that of the high-mobility component increases. This is expected, as the total resistivity (the fit to the experimental data) must remain the same.

%\clearpage

\section{Theory description}

\begin{figure}[!ht]
    \centering
    %\resizebox{1.3\textwidth}{!}{
    \includegraphics[width=\linewidth]{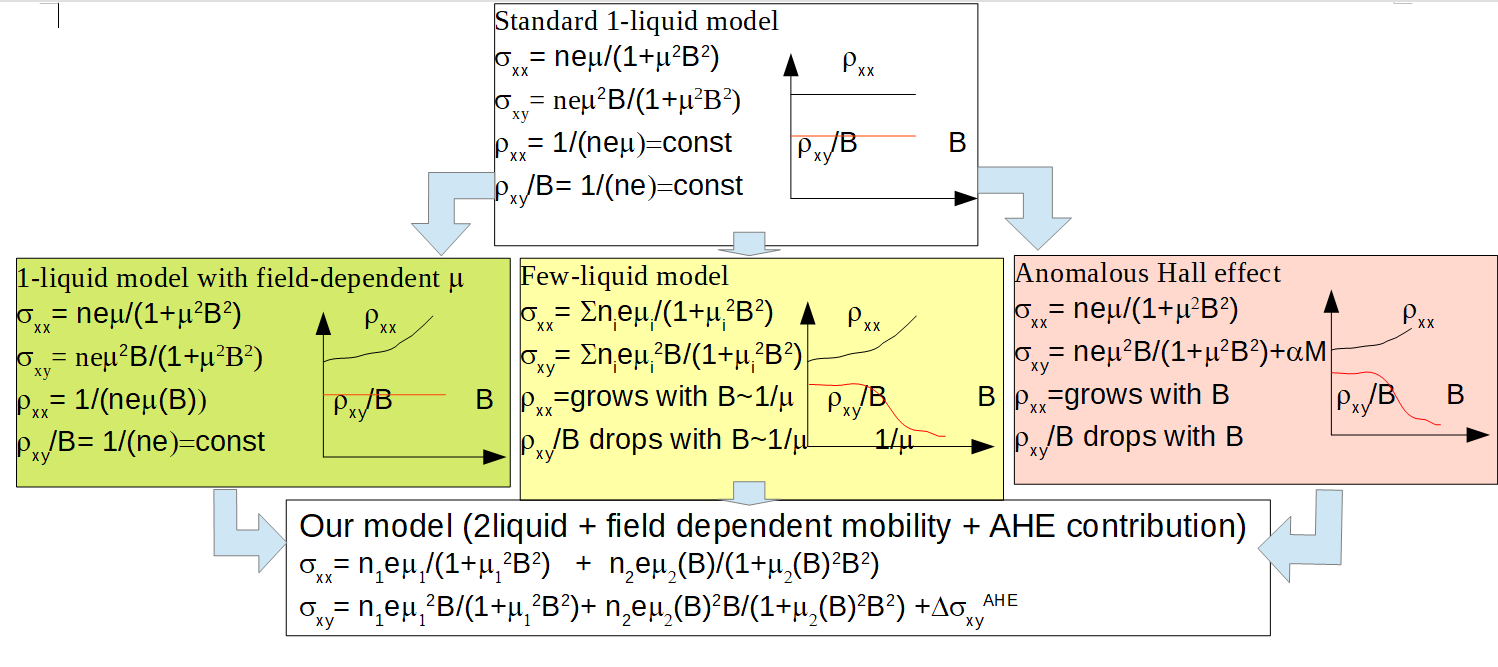}
    \caption{Interplay of various magnetoresistance and Hall effect nonlinearities mechanisms.}
    \label{theory}
\end{figure}

Let us consider the technical details of how magnetoresistance and Hall nonlinearity are formed within the relevant mechanisms, see Fig.\ref{theory}.

The conductivity tensor is characterized by two components ($\sigma_{xx}$ and $\sigma_{xy}$) in our case, due to the absence of ferromagnetism and in-plane anisotropy. It is explicitly recalculated into the experimentally measured resistivity tensor:

$$\rho_{xx}=\sigma_{xx}/(\sigma_{xx}^2+\sigma_{xy}^2)$$
$$\rho_{xy}=\sigma_{xy}/(\sigma_{xx}^2+\sigma_{xy}^2)$$
Within the standard classical one-band metal model, there should be NO magnetoresistance and a constant Hall coefficient (top bar in Fig.\ref{theory}), because the Lorentz force acting on electrons in a magnetic field $B$ is exactly compensated by the Hall electric field.

If there were only field-dependent mobility (green bar in in Fig.\ref{theory}), this would lead to magnetoresistance but would have no effect on the Hall effect, contrary to our experimental data, where the strongest Hall nonlinearity is observed.

The next mechanism is the two- (or multi-) liquid model (yellow bar in Fig.\ref{theory}). The idea is as follows: if there are several types of carriers with different mobilities, their conductivity tensors are summed up. However, after inversion of conductivity tensor into a resistivity tensor, the numerator and denominator no longer cancel each other. As a result, both $\rho_{xx}$ and $\rho_{xy}/B$ acquire corrections with the typical magnetic field scale $1/\mu$ where $\mu$ is the highest mobility.

This is exactly our 4-parameter fit. As seen from our data, this fit requires an excessively high mobility to account for the low-field dependence of the Hall effect. However, this high mobility fails to explain the data at higher magnetic fields. In other words, an essential ingredient is missing in the theory.

It is natural, therefore, to combine the field-dependent mobility and the two-liquid model. A minimal model is, for simplicity, one where only one component has field-dependent mobility. The Zeeman effect due to a perpendicular magnetic field naturally drives such a dependence because it leads to a magnetic topological insulator Hamiltonian [T.~Chiba \textit{et al.}, Physical Review B {\bf 95}, 094428 (2017)]. However, its manifestation in 3D topological insulators appears to be more complicated. First, it affects the $\sigma_{xx}$ correction (Eq.~(27) in the aforementioned paper):

\begin{equation}
    \sigma_{xx}=ne\mu\frac{1}{1+4\alpha^2B^2},
\end{equation}
where $e$ is the elementary charge and $h$ is Planck constant. Since this is a purely Zeeman effect without any orbital contribution, we believe that it purely renormalizes the mobility due to changes in the electron spectrum:

\begin{equation}
    \mu(B)=\frac{\mu(B=0)}{1+4\alpha^2B^2}.
    \label{mu2vsB}
\end{equation}

However, there is a second effect: Zeeman splitting due to the specific spectrum of a 3D TI leads to an anomalous Hall effect (AHE) contribution to $\sigma_{xy}$ (formula~(25) of Ref.~[T.~Chiba \textit{et al.}, Phys. Rev. B \textbf{95}, 094428 (2017)]); upon substituting
$\displaystyle \xi=\frac{\alpha}{\sqrt{1+\alpha^{2}B^{2}}}$, one obtains:

\begin{equation}
  \Delta\sigma_{xy}^{AHE}= -\,\frac{e^2}{h}\,\alpha B\,
  \sqrt{1+(\alpha B)^2}\,
  \frac{4\!\left[1+2(\alpha B)^2\right]}{\left[1+4(\alpha B)^2\right]^2}\,,
  \label{AHE_expr}
\end{equation}

At low magnetic fields, the AHE (pink bar in Fig.~\ref{theory}) affects mostly the Hall coefficient. However, when the anomalous Hall contribution becomes comparable to $\sigma_{xx}$, it also leads to magnetoresistance. We add the AHE contribution to the Drude Hall component of the conductivity tensor.

Since the Zeeman effect has nothing to do with orbital motion, we believe that its effect on conductivity reduces to the renormalization of mobility $\mu(B)$ and the addition of the AHE. Orbital motion (the Lorentz force) comes into play through the $1/(1+\mu^2B^2)$ factor in the Drude conductivity tensor components.

Taking all these factors into account, we obtain the theoretical expressions for the conductivity tensor components shown in the bottom bar of Fig.~\ref{theory}, with $\Delta\sigma_{xy}^{AHE}$ given by Eq.~\ref{AHE_expr} and $\mu_1(B)$ given by Eq.~\ref{mu2vsB}.

Notably, only five parameters are sufficient to parametrize the magnetotransport within this theoretical approach: $n_1$, $n_2$, $\mu_1^0$, $\mu_2$, and $\alpha$. To extract all parameters, we fit $\rho_{xx}(B)$ and $\rho_{xy}(B)/B$ simultaneously.

%\end{document}

\end{document}